\documentclass{cfdsc}
\usepackage[T1]{fontenc}
\usepackage{setspace}
\usepackage{amssymb}

\usepackage{multicol,caption}
\usepackage{graphicx,bm,bigints}

\usepackage[square,numbers]{natbib}

\newenvironment{Figure}
  {\par\medskip\noindent\minipage{\linewidth}}
  {\endminipage\par\medskip}

\begin{document}

\title{Low-dimensional representation of fluid flows using proper orthogonal decomposition}


\author{Jahrul Alam and Asokan Variyath}

\institute{Department of Mathematics and Statistics, Memorial University of Newfoundland, \\ St John's, Canada }

\email{alamj@mun.ca}

\begingroup
\onecolumn{
{\begin{flushright} \small{{Proceedings of the 29th Annual Conference of the Computational Fluid Dynamics Society of Canada}\\{CFDSC2021}\\{July 27-29, 2021, St. John's, NL, Canada}}
\end{flushright} }}

\maketitle

\endgroup


\begin{multicols}{2}

  \begin{abstract}
    The fluid flow around a bluff body is complex and time dependent, which also contains a wide range of time and length scales.  The first few eigenmodes of the proper orthogonal decomposition (POD) of such a flow provide significant insight into the flow structure, and can form the basis of a low-dimensional representation of certain turbulent flows. In this article, the direct-forcing immersed boundary method is considered to model the wake flow generated by arbitrary shaped obstacles. Based on the POD analysis of wakes behind cylinders, airfoils, and rotors, a reduced order model (ROM) for the prediction of wake dynamics is studied for arbitrary solid obstacles. For low Reynolds number time periodic flows, the POD based ROM accurately captures the statistically representative coherent motion. For high Reynolds number atmospheric boundary layer flow around rotors, POD provides a low-dimensional representation of the meaningful statistics of coherent motion. The POD based ROM is analyzed for turbulent flow past a rotor in the atmospheric boundary layer, where the flow is not periodic in time.

  \end{abstract}

%

\section{Background and objective}
%

The proper orthogonal decomposition~(POD) was introduced by Lumley~\cite{Lumley67} for the analysis and detection of coherent structures in turbulent flows. It is a statistical learning method that extracts spatial and temporal behaviors of coherent structures in a fluid flow~\cite{Cazemier98}. The performance of the POD  method in detecting coherent structures in various turbulent flow regimes is is an active area of research~\cite{Xue2020}\cite{Lim2020}~\cite{Ping2020}. POD forms a basis for the reduced order model (ROM) to represent the Navier-Stokes dynamical system, where a few leading POD modes can reproduce the coherent part of a turbulent flow~\cite{Abbaszdeh2021}. POD is similar to principal component analysis~(PCA)~\cite{James2013}\cite{Rahoma2021}\cite{Lim2021}, or empirical orthogonal functions~\cite{Lumley67}. It has been successfully applied to image processing, pattern recognition, and in many other engineering and machine learning applications~(see~\cite{Luo2018}).

The objective of this article is to explore and evaluate the POD method, where the immersed boundary method (IBM)is  combined with large eddy simulation (LES) of fluid flows around complex geometries. The main motivation of the research on POD is twofold: to gain a better understanding of the modal decomposition methods and to design effective large eddy simulation (LES) methods for fluid-solid interaction problems. Such a development might be useful for the design, control and optimization of wind power plants and similar applications~\cite{Marek2021}\cite{Debnath2017}\cite{Bi2021}. The  ultimate  goal  of  our  research  includes the  reduction of computational complexity in LES of turbulence around arbitrary objects, such as spheres, cylinders, airfoil etc, as well as turbulence within wind farms so that a prescribed accuracy is achieved with  limited  computational costs. 

POD offers a scale separation similar to that of LES~\cite{Shinde2020}. However, the number of computational degrees of freedom in POD based ROM is much less than that of LES. If a turbulent flow is not periodic in time, the POD modes do not represent actual instantaneous coherent structures of turbulence. Nevertheless, the results of POD based ROM in cylinder wakes are usually in agreement with that of LES, at least for short-time integration~\cite{Marek2021}. In the POD analysis of wake behind a circular cylinder, it is generally observed that the onset of divergence from the periodic vortex shedding depends on the number of POD modes retained for the approximation, the Reynolds number, and the flow geometry. Striking a balance between the efficiency and the accuracy in ROMs of turbulent flows remains one of the main challenges.  

In fluid-solid interaction scenario, there exists local regions where pressure may be decreased or increased, thereby leading to local regions containing a favorable or adversed pressure gradient. In case of adversed pressure gradient, the flow decelerates, thickens, and may develop a point of inflection in the vicinity of solid boundaries. The adversed pressure gradient can reverse the direction of flow. Experimental observations of such fluid flows suggest that the trailing surface of a slender body may overcome the impact of pressure drag penalty, suggesting to gradually reduce the size of the solid body in order to control the near-wake dynamics. It is our intention to utilize such a knowledge of dynamics in order to accurately model the underlying of fluid-solid interaction phenomena in LES, while employing a low-cost Cartesian mesh~\cite{Alam2018}\cite{Alam2018a}\cite{Alam2020}. Hence, we need to understand the characteristics of coherent structures captured with statistical learning tools, such as the POD or dynamic mode decomposition (DMD) based reduced order modeling approaches~\cite{James2013,Debnath2017}. For this purpose, first, we perform POD of the turbulent flow into linearly uncorrelated scales of turbulence. Second, we compare the subfilter-scale motion between LES and in terms of the POD modes.

In Section~\ref{sec:meth}, we discuss the computation of POD modes through the PCA approach, as well as the Galerkin projection of the Navier-Stokes equation on to the POD modes. In Section~\ref{sec:res}, we discuss the POD analysis for time periodic fluid flow and atmospheric boundary layer flow around rotors. Finally, we summarize the findings of the present study in Section~\ref{sec:con}.  




\section{Methodology}\label{sec:meth}
\subsection{Proper orthogonal decomposition}
To decompose a velocity field in POD modes, consider a snapshot of the fluctuating part of the velocity field $\mathcal X = \{u'_i(\bm x_k,t^n)$ for $i=1,2,3$, $k=1,2\ldots,\mathcal N$, and $n=1,2,\ldots,M\}$. For all grid points $\{\bm x_k\}$, let us stack the velocity at $n$-th column of a matrix $\mathcal X$ as $\bm u^n = [u'_1(\bm x_1,t^n),u'_2(\bm x_1,t^n),u'_3(\bm x_1,t^n),\cdots,]^T$. It maybe convenient to denote each column of the snapshot matrix $\mathcal X$ by  $\bm u^n = [u'_j(t^n)]$, where $j=1,\ldots,3\mathcal N$. PCA solves the following $M\times M$ eigenvalue problem
\begin{equation}
  \label{eq:Ai}
  \left[\frac1M\mathcal X^T\mathcal X\right]\mathcal A_n = \sigma_n\mathcal A_n,\quad n=1,\ldots,M,
\end{equation}
where $\mathcal A_n\in\mathbb R^M$ is the eigenvector, which represents the temporal variability in $\mathcal X$~\cite{Marek2021}. The eigenvalues $\sigma_n$ denote the energy contained in PCA modes.

In the POD approach, the fluctuations in the fluid velocity is decomposed in space and time, 
\begin{equation}
  \label{eq:pod}
  \bm u'(\bm x,t) = \sum_{j=1}^{r}a_j(t)\bm \varphi_j(\bm x),
\end{equation}
where $\bm\varphi_j(\bm x)\in\mathbb R^3$ is continuous vector-valued function, and $\{a_j(t)\}$ denotes a continuous function corresponding to each component of $\bm u'$. In the present work, we consider the Reynolds decomposition $\bm u(\bm x,t) = \bar{\bm u}(\bm x) + \bm u'(\bm x,t)$, where $\bm u(\bm x,t)$ is  the velocity field resolved with LES.

%
Eq~(\ref{eq:pod}) is a projection of $\bm u'$ on to a $r$-dimensional subspace ($r\ll M$) spanned by the set $\{\bm\varphi_j(\bm x)\}_{j=1}^r$ such that the difference between $\bm u'$ and its low-dimensional projection onto $\{\bm\varphi_j(\bm x)\}_{j=1}^r$ is minimized by solving a constrained optimization problem, which is reduced to
\begin{equation}
  \label{eq:pca}
  \sum_{j=1}^M\mathcal R_{ij}\bm\varphi_j = \sigma_i\bm\varphi_i, \quad i=1,\ldots M
\end{equation}
where $[\mathcal R_{ij}]\equiv \langle \bm u'(\bm x,t^i)\otimes\bm u'(\bm x,t^j)\rangle$. 
Each element of the $M\times M$ matrix $[\mathcal R_{ij}]$ is a continuous function of $\bm x$, where
the operator $\mathcal R$ is self-adjoint and positive semi-definite. Thus, there exists a countable set of non-negative eigenvalues $\sigma_i$ associated with eigenfunctions $\bm\varphi_i(\bm x)$. Solving Eq~(\ref{eq:pca}) on a set of $\mathcal N$ grid point $\bm x_k$, we can find $\bm\varphi_i(\bm x_k)$, which consists of solving $\mathcal N$ eigenvalue problems in order to determine the discrete POD modes. Clearly, finding the POD modes is computationally challenging task.

In practice, Eq~(\ref{eq:pod}) is discretized in space and time. Thus, for each $j$ the discrete time coefficients $a_j(t_n)$, $n=1,\ldots,M$ of the discretized POD Eq~(\ref{eq:Ai}) can be taken the same as the eigenvectors in the PCA Eq~(\ref{eq:pca}). In other words, the PCA problem (\ref{eq:Ai}) can be solved once in order to find the PCA vectors $\mathcal A_n$. The discrete POD basis functions are computed for $n=1,\ldots,M$ as
$$
[\bm\varphi_n(\bm x_k)] = \mathcal X\mathcal A_n.
$$

\subsection{Reduced order model for Navier-Stokes equation}
%

A reduced order model of fluid flow is derived from the POD of the velocity field. We consider the Galerkin projection of the Navier-Stokes equation onto the POD basis $\{\bm\varphi_j\}$. Substituting
\begin{equation}
  \label{eq:ak}
        {\bm u}(\bm x,t) = \bar{\bm u}(\bm x) + \sum\limits_{k=1}^ra_k(t)\bm\varphi_k(\bm x)
\end{equation}
into the incompressible Navier-Stokes equation and applying the orthonormality property of the basis functions $\{\bm\varphi_j\}$ gives~\cite{Akhtar2009}\cite{Luo2018},
\begin{equation}
  \label{eq:rom}
\frac{da_k(t)}{dt} = \mathcal M_k + \sum_{m=1}^M\mathcal L_{km}a_m(t) + \sum_{m=1}^M\sum_{n=1}^M\mathcal C_{kmj}a_n(t),
\end{equation}
where
$$
\mathcal M_k = \frac{1}{\nu}\langle\bm\varphi_k,\nabla^2\bar{\bm u}\rangle - \langle\bm\varphi_k,\bar{\bm u}\cdot\bm\nabla\bar{\bm u}\rangle
$$
$$
\mathcal L_{km} = -\langle\bm\varphi_k,\bar{\bm u}\cdot\bm\nabla\bm\varphi_k\rangle - \langle\bm\varphi_k,\bm\varphi_m\cdot\bm\bar{\bm u}\rangle + \frac{1}{\nu}\langle\bm\varphi_k,\nabla^2\bm\varphi_m\rangle
$$
$$
\mathcal C_{kmn} = -\langle\bm\varphi_k,\bm\varphi_m\cdot\bm\nabla\bm\varphi_n\rangle,
$$
and $\langle \bm a,\bm b\rangle = \int\bm a\cdot\bm b\,d\bm x$ denotes the usual inner product. Note that the Green's theorem and the divergence-free property allows to drop the pressure gradient term from the ROM equations~\cite{Akhtar2009}.

\section{Results and discussion}\label{sec:res}
In the following analysis, we use a LES code to solve the filtered incompressible Navier-Stokes equation~\cite{Alam2018}. To analyze the POD based low-dimensional representation of fluid flow around obstacles, snapshots of the velocity field have been analyzed for three canonical cases: a circular cylinder, an airfoil, and a rotor disk. A goal of this analysis is to understand how to address the challenges of turbulence in the modern field of data-driven discovery of ROMs for turbulent flows. Here, the Reynolds number $\mathcal Re=U_\infty D/\nu$ is obtained using the free stream velocity $U_\infty$, the characteristic size  $D$ of the solid obstacle, and the kinematic viscosity $\nu$ of air. The canopy stress method discussed in~\cite{Alam2018} is considered to implement the direct-forcing IBM. 

The domain for the flow past a cylinder is $32D\times 3D\times 6D$, the inlet boundary is located at a distance of $6D$ from the center of the cylinder, and the cylinder is parallel to the $y$-axis~\cite{Ping2020}\cite{Marek2021}. The number of cells are $576\times 54\times 108$ in the $x$-, $y$-, and $z$-directions, respectively. There are about $20$ cells across the diameter of the cylinder. In the POD analysis discussed below, a snapshot matrix of the 3D velocity of size $10\,077\,696\times 421$ was collected for the time period $t\in[89,110]$. Similarly, Reynolds number and the computational domain for the simulation of flow around an airfoil are kept the same as that of the cylinder case. 

In case of the flow past a rotor, we have considered a computational domain of size $36D \times 6D\times 3D$ and a mesh with $256\times 36 \times 72$ cells  in the stream-wise ($x$), span-wise ($y$), and vertical directions ($z$), respectively. There are about $7$ cells across the rotor. The Reynolds number is $\mathcal Re=4\times 10^5$ and $D=0.15$~m. The data from an wind-tunnel experiment~\cite{Wu2011} is considered to simulate the atmospheric boundary layer flow around a single rotor.

\subsection{Flow past a cylinder}
The criterion for identifying a instantaneous coherent vortices may simply be based on filtering the turbulent flow, where coherent vortices are what is left after turbulence has been denoised. Instantaneous coherent vortices can be identified by the eigenvalues of the symmetric tensor $\mathcal S^2+\mathcal R^2$ ($\lambda_2$ criterion) or by $(1/2)(||\mathcal R||^2-||\mathcal S||^2)$ (Q-criterion). In the cylinder flow, $y$-component of the vorticity $\omega_2=\epsilon_{2jk}\partial u_k/\partial x_j$ is commonly used for the visualization of coherent flow structures, where $\epsilon_{ijk}$ is the alternating tensor. 

Fig~\ref{fig:sig}$a$ shows the normalized energy distribution of $421$ POD modes for the cylinder flow at $\mathcal Re = 200$, where the snapshots are taken from the time-span of $89$~s to $110$~s. The cumulative energy of the first $50$ POD modes are displayed in Fig~\ref{fig:sig}$b$. It is clear that about $10$ of the POD modes contain the most of the energy. 
\begin{Figure}
  \centering
  \begin{tabular}{c}
    \includegraphics[height=6cm]{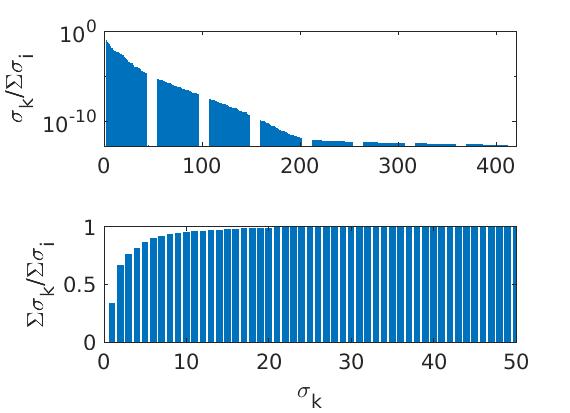}
  \end{tabular}
  \captionof{figure}{(a) Normalized energy contribution from individual POD modes (top) and (b) normalized cumulative energy of POD modes (bottom).}
  \label{fig:sig}
\end{Figure}

In Fig~\ref{fig:rom}, we visualize the vortex shedding pattern in the 3D velocity field around a circular cylinder at $t=110$~s, where the velocity is restricted on the 2D vertical mid-plane of the domain and the vorticity is calculated from the restricted data. In comparison to the simulation of ref~\cite{Marek2021}, using a similar numerical configuration, it is observed that the vortex shedding phenomena has been captured reasonably accurately with our LES-IBM code, which is sufficient for the purpose of present modal analysis study.

Instantaneous coherent vortices directly captured with LES for $t=110$~s is compared ({\em e.g.} Fig~\ref{fig:rom}$a$)  with the coherent vortices captured statistically with the most dominant $10$ of the $421$ POD modes (Fig~\ref{fig:rom}$b$). Notice that the vorticity field $\omega_2$ for $t=110$~s in Fig~\ref{fig:rom}$b$, which is reconstructed with only $10$ POD modes, compares reasonably well with that in the vorticity of the original data shown in Fig~\ref{fig:rom}$a$. From Fig~\ref{fig:sig}$b$, we see that a majority of the energy is contained in the strongest $10$ POD modes.

Fig~\ref{fig:rom}$c$ shows the vorticity $\omega_2$ at $t=110$~s, which is predicted with the ROM using the $10$ dominant eigenmodes of the principal components. More specifically, the Galerkin project of the Navier-Stokes equations onto the strongest $10$ POD modes provides the POD expansion coefficients $a_n(t)$ at $t=89$~s. Solving the reduced system, Eq~(\ref{eq:rom}), provides $a_n(t)$ at $t=110$~s, and hence, $\omega_2$ at $t=110$~s. The vortex-shedding pattern for the low-dimensional representation of $\bm u(\bm x,t)$ at $t=110$~s in Fig~\ref{fig:rom}$b$ and Fig~\ref{fig:rom}$c$ compare well with the corresponding LES result in Fig~\ref{fig:rom}$a$. In Fig~\ref{fig:romu}, we present $\omega_2$ restricted on a line in the vertical mid-plane of the domain passing through the center of the cylinder. Running ROM with $10$ and $50$ pod modes and comparing with the computational fluid dynamics (CFD) results, we notice that the result of ROM is not highly sensitive to the number of POD modes, as long as the most of the energy is captured by the POD modes.
%
%
%
%
%
 \begin{Figure}
  \centering
   \begin{tabular}{c}
     $(a)$\\
     \includegraphics[height=2.5cm]{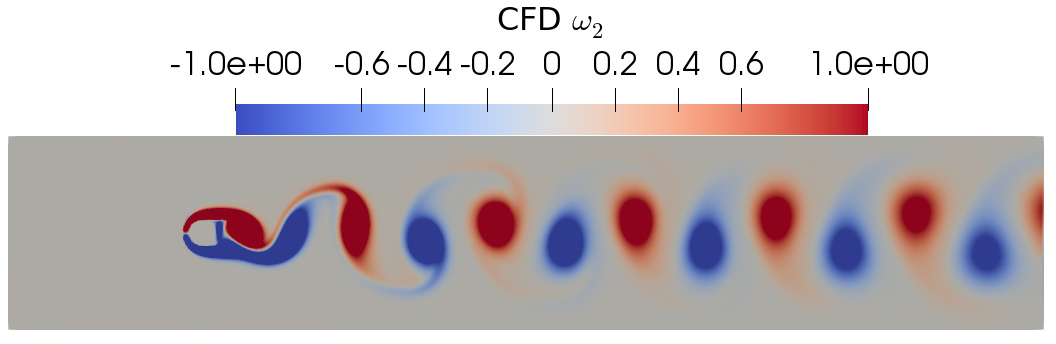}
     \\
     $(b)$\\
     \includegraphics[height=2.5cm]{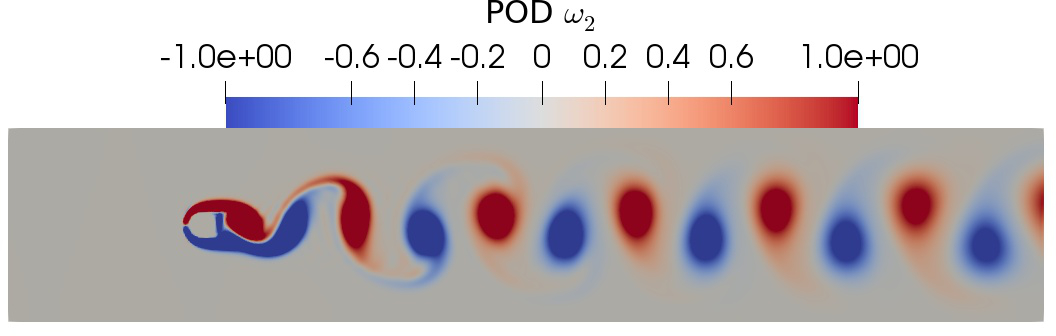}
     \\
     $(c)$\\
     \includegraphics[height=2.5cm]{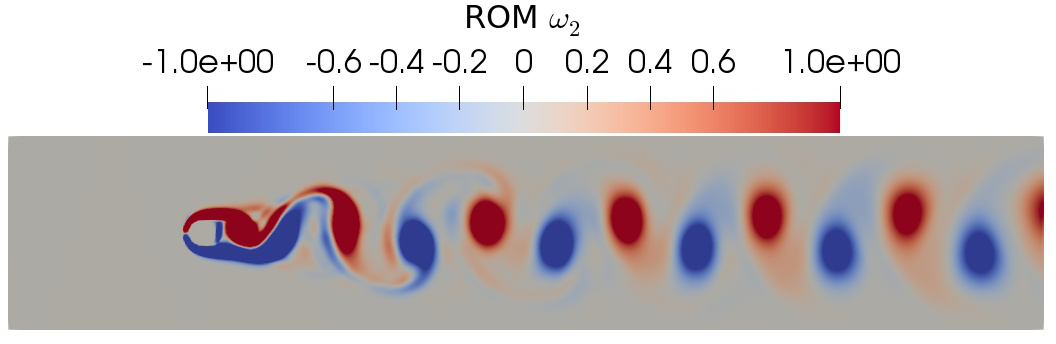}
   \end{tabular}
   \captionof{figure}{A comparison of the $y$-component vorticity $\omega_w$ for the wake behind a cylinder at $t=110$~s. }
   \label{fig:rom}
 \end{Figure}
 \begin{Figure}
  \centering
   \begin{tabular}{c}
     \includegraphics[width=7.5cm]{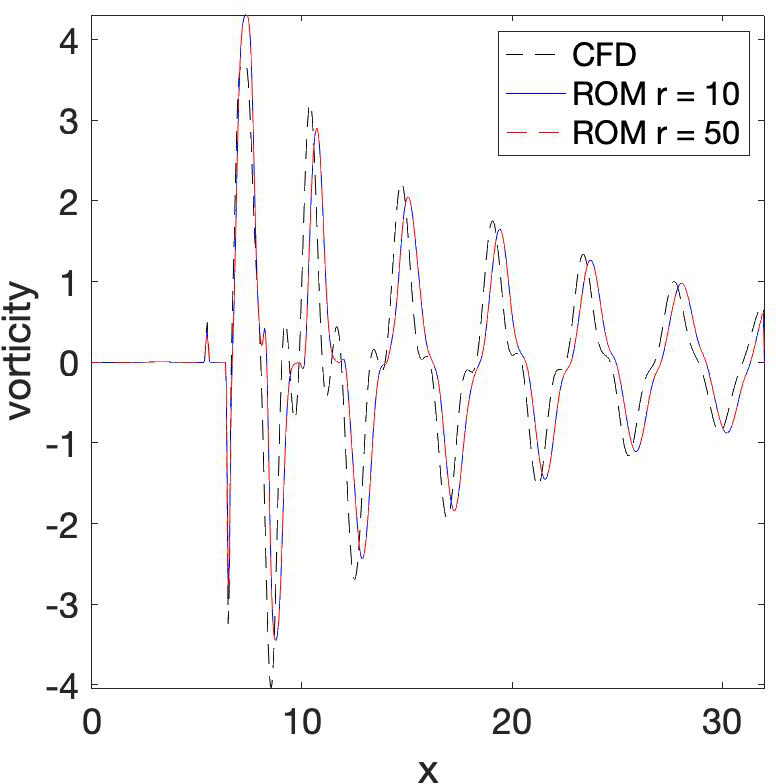}
   \end{tabular}
   \captionof{figure}{The periodic fluctuation of the vorticity in the wake behind a cylinder at $\mathcal Re = 200$.}
   \label{fig:romu}
 \end{Figure}
 In a similar manner to the velocity snapshots considered here, the snapshots of the vorticity field, $\omega=\bm\nabla\times\bm u$, can be considered in the POD analysis ({\em e.g.} see~\cite{Kutz2016}). The eigenmodes of the vorticity snapshot would capture the enstrophy (mean squared vorticity) of the flow, instead of energy. In the two-dimensional turbulence theory, both the energy and the enstrophy are conserved. However, enstrophy is not conserved in 3D turbulence due to vortex stretching. The POD modes of either velocity or vorticity provides the insight into statistically representative coherent structures, meaning that temporal fluctuations over mean flow can be represented into vortices identified by classical methods.

 The POD method has received historical recognition in fluid dynamics community, in order for representing turbulence data with minimal number of basis functions or modes while preserving as much energy as possible~\cite{Zhang2020}\cite{Katzenmeier2020}\cite{Krath2021}. In POD based ROM, a mathematical measure of the reducibility is defined as
 \begin{equation}
   \label{eq:kwth}
   d_n(\mathcal X) := \inf\,\sup||\bm u -\mathcal P\bm u||
 \end{equation}
 where the infimum ($\inf$) is taken over all possible $n$-dimensional  subspace, $\bm u$ is the state on the solution manifold, and $\mathcal P$ is the orthogonal projector onto the space of POD modes. Based on the decay rate of $d_n(\mathcal X)$ with increasing $n$, the reducibility and the quality of reduced order approximation can be judged. The slow decay of the Kolmogorov width in turbulent flows is often denoted as the ``Kolmogorov barrier''~\cite{Alam2007}. It has been estimated that the number of spatial modes required to uniquely determine a two-dimensional turbulent flow is bounded by $\mathcal Re^{4/3}$ and $\mathcal Re$, respectively, for forced and decaying turbulence~\cite{Davidson2004}. At fixed $\mathcal Re=200$, the flow around a 3D cylinder possesses a two-dimensional dynamics. The barrier in the decay of Kolmogorov width is clearly depicted from Fig~\ref{fig:romu}, where it is seen that the vortex-shedding pattern obtained from the ROM simulation is not sensitive to the number of POD modes. A comparison of the vorticity at $t=110$~s between the ROM simulation using $10$ POD modes and that for $50$ POD modes is shown in Fig~\ref{fig:romu}, which does indicate a rapid convergence toward true solution. It may also be noted from Fig~\ref{fig:romu} that the vortex-shedding in the ROM simulation (Fig~\ref{fig:rom}$c$) exhibits  dispersive errors in comparison to the CFD results (Fig~\ref{fig:rom}$a$). Other machine learning approaches, such as a long short-term memory nudging algorithm may be incorporated to further enhance the performance of ROM in simulating high Reynolds number fluid flows~\cite{Brunton2020}\cite{Kutz2016}.

 \subsection{Flow past an airfoil}
 Consider the flow around an airfoil at $\mathcal Re = 200$. Unlike cylinders, airfoils are streamlined bodies. The goal is to understand the effects of changing the shape of the solid body, without changing other parameters with respect to the cylinder flow~\cite{Zhu2020}.  Fig~\ref{fig:wake}$a$ shows the vorticity $\omega_2$ at $t=110$~s. It can be seen that the vortex shedding behind an airfoil (Fig~\ref{fig:wake}$a$) is different than that is shown in Fig~\ref{fig:rom} for a cylinder. Nevertheless, both flow becomes periodic in time. It is usually observed for such a time periodic flow that only a few POD modes provide significant insight into the flow structure, and can form the basis of a reduced dynamical model.

 To understand how the POD method captures the statistically representative coherent structures in a fluid flow, we can visualize  the POD basis $\bm\varphi_i(\bm x)$. In Fig~\ref{fig:wake}$b$ and Fig~\ref{fig:wake}$c$, we present the color-filled contour plots of the 1st POD mode for the airfoil case, corresponding to the $x$-and $z$-components of the velocity $\bm u(\bm x,t)$, respectively, where the number of snapshots $421$ is kept the same as what is used for the cylinder flow. Fig~\ref{fig:wake}$b$ indicates that the strongest POD mode of the $x$-component velocity signifies the vortex-shedding pattern. Similar results are usually observed in cylinder flow (see~\cite{Marek2021}). 
The brief analysis presented in this section, we show the ability of the POD method to extract coherent structures in fluid flows around streamlined bodies~\cite{QZhang2014}. Although a relatively comprehensive investigation is needed to fully understand POD assisted ROM of wakes behind obstacles, particularly of arbitrary shapes, the present brief study indicates that POD method works well for the time-resolved wake flow data in extracting the statistics of coherent structures. It is worth mentioning that the POD method presented in this article is still limited to low Reynolds number time-periodic flows.
 \begin{Figure}
  \centering
   \begin{tabular}{c}
     $(a)$\\
     \includegraphics[height=2.5cm]{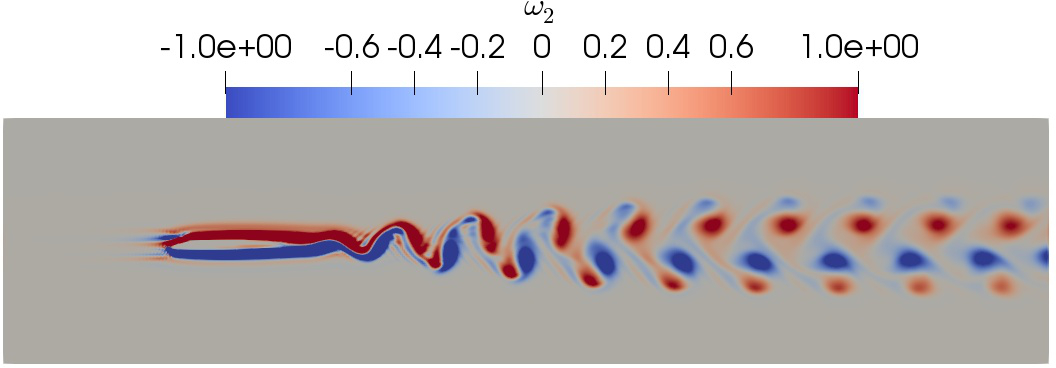}
     \\
     $(b)$\\
     \includegraphics[height=2.5cm]{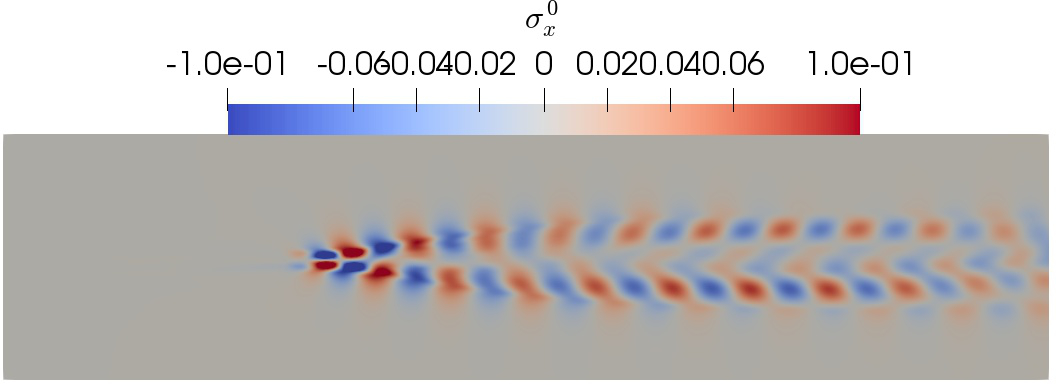}
     \\
     $(c)$\\
     \includegraphics[height=2.5cm]{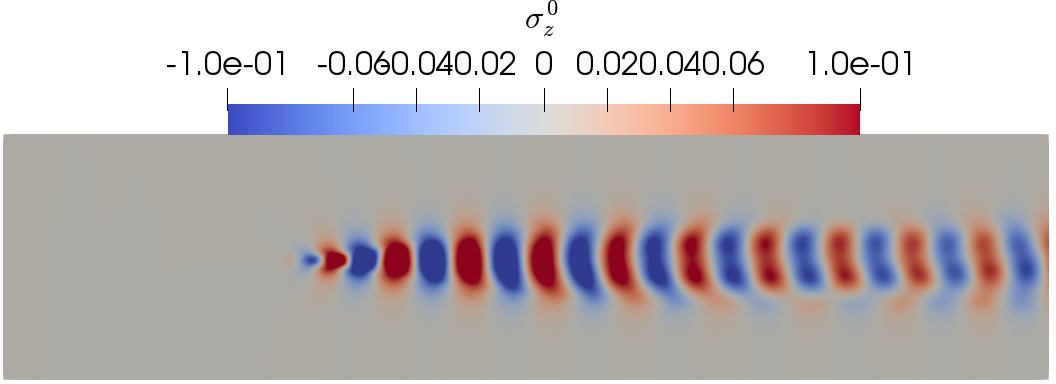}
   \end{tabular}
   \captionof{figure}{The wake behind an airfoil. $a$ The $y$-component vorticity $\omega_2$. $b$ The first POD mode $\sigma_x^0(x,z)$ corresponding to the $x$-component velocity, which is projected on the vertical mid-plane of the domain. $c$ The first POD model $\sigma_z^0(x,z)$ corresponding the $z$-component velocity, which is projected on the vertical mid-plane. }
   \label{fig:wake}
 \end{Figure}
 \subsection{A rotor in the atmospheric boundary layer}
\begin{Figure}
  \centering
   \begin{tabular}{cc}
     \includegraphics[height=5.0cm]{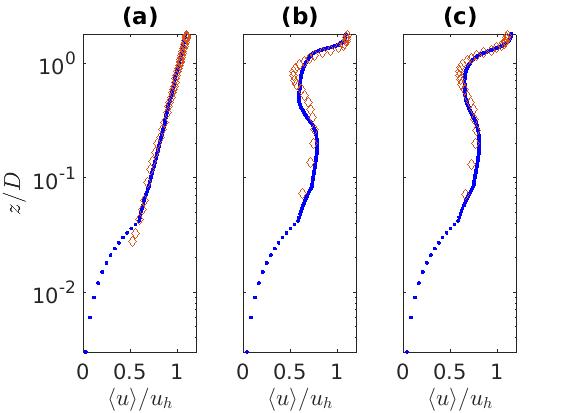}
     &
     \includegraphics[height=5.0cm]{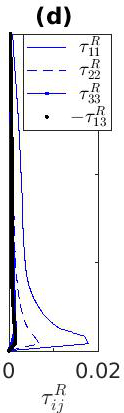}
   \end{tabular}
   \captionof{figure}{($a-c$) A comparison of $\langle u(z)\rangle/u_h$ between LES and wind tunnel measurements, where $u_h$ is the reference wind speed at hub height in the inlet boundary. $(a)$~At the inlet, $(b)$ at a downstream distance of $2D$ from the rotor, $(c)$ at a downstream distance of $3D$ from the rotor, and $(d)$ Distribution of Reynolds stresses, $\tau_{ij}^R$ at a downstream distance of $3D$ from the rotor. }
   \label{fig:adm}
 \end{Figure}

 Unlike a cylinder or an airfoil, where periodic vortex shedding is a primary pattern of the dynamics, the flow around a rotor operating in the atmospheric boundary layer is relatively complex. Moreover, the genesis of coherent vortices in the wakes of a cylinder, an airfoil, or a rotor is an effect the viscous stresses leading to flow structures associated with the vortex filaments. However, POD method would capture statistically representative coherent flow structures that are dynamically different than the instantaneous coherent vortices detected by other methods, such as $\lambda_2$-, $\Delta$-, or Q-criterion. In this view, we now focus on understanding the POD method for atmospheric boundary layer flow around a rotor.

Consider a single rotor with diameter $D$ and hub height $z_h = 0.83D$, which is placed at a downstream distance of $4.5D$ from the inflow boundary of a computational domain of size $36D \times 6D\times 3D$. We have used a mesh with $256\times 36 \times 72$ cells  in the stream-wise, span-wise, and vertical directions, respectively.  The IBM approach is employed to model the thrust force of the rotor, which is a modification of the classical actuator disk model of wind turbines.  For $D=0.15$~m, there are about $7$ cells across the rotor. This resolution is about $50$\% coarser than $10$-$15$ cells  across the rotor, which is usually required in LES of flow around actuator disks. The Reynolds number is $\mathcal Re=4\times 10^5$ for this simulation. In the present LES, we match the mean stream-wise velocity $\langle u(z)\rangle = (u_*/\kappa)\ln(z/z_0)$ at the inflow boundary with that from the wind tunnel experiment (see~\cite{Stevens2018}). To simulate coherent structures associated with dynamic variations of wind direction in atmospheric boundary layers, we consider random perturbations to the span-wise and vertical components of the velocity in the inflow boundary. 

It was generally observed from measurements of wind and turbulence in actual wind farms that wind turbines act as an elevated sink of momentum and a source of turbulence kinetic energy~\cite{Roy2011}. Here, the direct forcing IBM approach models the axial thrust imparted onto an wind turbine, which is very similar to the actuator disk model. 

In Fig~\ref{fig:adm}$a$, the normalized stream-wise velocity at the inflow boundary is compared with that from experiments. The average momentum deficit in the wake behind a rotor disk is presented in Fig~\ref{fig:adm}$b$-$c$, where the stream wise velocity $\langle u(z)\rangle/u_h$ from LES is also compared with that from wind tunnel measurement~\cite{Stevens2018}.  With respect to two downstream locations, which are at distances of $2D$ and $3D$, respectively, from the rotor, the axial thrust predicted with the IBM approach indicates a good agreement with that of the wind tunnel measurements.

  \begin{Figure}
  \centering
   \begin{tabular}{c}
     $(a)$\\
     \includegraphics[height=2.35cm]{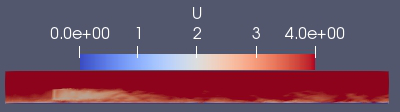}
     \\
     $(b)$\\
     \includegraphics[height=2.35cm]{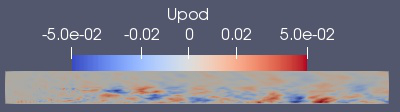}
     \\
     $(c)$\\
     \includegraphics[height=1.8cm]{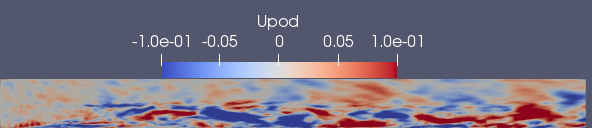}
     \\
   \end{tabular}
   \captionof{figure}{A contour plot of the stream-wise velocity on the vertical mid-plane of the domain. $(a)$ Mean stream-wise velocity $\langle u(x,y_c,z)\rangle$ with respect to $751$ snapshots. $(b)$ Perturbation to stream-wise velocity based on $2$ POD modes. $(c)$ The same as $(b)$ for $4$ POD modes. }
   \label{fig:pod}
 \end{Figure}

Turbulence intensity has been computed from snapshots of velocity during a period of $75$ seconds, which consists of about $126$ eddy turn over time units. The duration is reasonable to accurately capture the Reynolds stress $\tau^R_{ij}$ {\em via} the covariant tensor of fluctuating velocity around the rotor. The diagonal elements of $\tau^R_{ij}$ represents turbulence intensity, which are the variances of the fluctuating velocity. Fig~\ref{fig:adm}$d$ presents the diagonal components of the Reynolds stress $\tau_{ij}^R$, as well as one of the off-diagonal components, $-\tau_{13}^R$, that represents the shear stress experienced by the ground boundary. Similar to high Reynolds number turbulent boundary layers, Fig~\ref{fig:adm}$d$ shows that turbulence intensity is about $20$\% of the inflow velocity. Notice that the LES of single rotor has approximately captured the known dynamics of atmospheric boundary layer flow around a single wind turbine. 

  There exists a relatively little investigation on the POD method pertaining to the detailed dynamics of atmospheric boundary layer turbulence, where an array of wind turbines causes momentum deficit while increasing turbulence intensity. The performance of a wind turbine operating in the wake behind another turbine will be affected by the  dynamics  of  the  velocity  field  resulting  from the  interaction  between  the  atmospheric  boundary layer and the upstream wind turbine. To advance the POD method in this direction, a finite number of point vortices are randomly injected on the inlet plane, where the corresponding velocity $[0, v', w']^T$ is obtained by using the Biot-Savart law. No perturbation is injected into the stream-wise velocity to ensure that LES retains a mean flow which is similar to that of the wind tunnel experiment. For any random vector field $[\bm\omega'(\bm x'),0,0]^T$ injected at random locations $\bm x'$ in the inlet plane,  the Biot-Savart law provides the induced velocity perturbation at $\bm x$, where
  $$
  \bm u'(\bm x) = -\frac{1}{4\pi} \int\frac{\omega'(\bm x')\times(\bm x-\bm x')}{|\bm x-\bm x'|^3}d^3\bm x'.
  $$
  In other words, the direction of inlet velocity is dynamically altered without altering the mean atmospheric boundary layer profile. The wind direction in atmospheric boundary layers varies continuously. Using POD analysis, we show that dynamic variations of wind direction may affect (either positively or negatively) the performance of large wind farms.

  \begin{Figure}
    \centering
    \begin{tabular}{c}
      $(a)$\\
      \includegraphics[height=2.35cm]{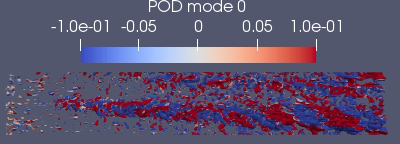}
      \\
      $(b)$\\
      \includegraphics[height=2.40cm]{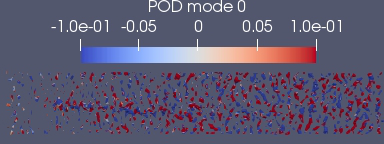}
      \\
    \end{tabular}
    \captionof{figure}{Q-criteria in the wake behind a single rotor in the atmospheric boundary layer. $(a)$ Q-criteria (Q=150) was constructed from $4$ POD modes - the corresponding velocity is shown in Fig~\ref{fig:pod}$c$. $(b)$ The Q-criteria of the original velocity field obtained with LES. Color-filled contour plot of the first POD mode is also overlayed with each plot.}
    \label{fig:podQ}
  \end{Figure}

  Fig~\ref{fig:pod}$a$ displays a contour plot of the mean stream-wise velocity $U(x,y_c,z)$ which is an average of $751$ snapshots of the instantaneous velocity $u'(x,y,z,t)$ at time steps of $0.1$~sec. During the period of $75$ seconds considered here, the mean velocity was found to reach an approximate steady state, and the LES results also show a good agreement with experimental data (see Fig~\ref{fig:adm}). The spatial fluctuations with respect to the mean stream-wise velocity are displayed in Fig~\ref{fig:pod}$b$ and Fig~\ref{fig:pod}$c$, where $u'(x,y_c,z)$ is reconstructed using only $2$ and $4$ POD modes, respectively. Here, $y_c$ indicates that the vertical restricted on the vertical midplane. It can be seen that the POD modes are useful tool to understand wake behind a rotor. Fig~\ref{fig:pod}$a$ shows that the wind has approximately recovered to the inlet profile at a downstream distance of $8D$-$10D$ from the rotor while turbulence stresses may be confined in the vicinity of the ground below the level of the rotor. However, Fig~\ref{fig:pod}$c$ indicates that the coherent structures of the strongest POD modes acts in entraining kinetic energy from aloft~\cite{calaf2010}.

  Fig~\ref{fig:podQ} shows compares the Q-criteria between the velocity field of LES and that reconstructed with $4$ POD modes. The Q-criterion is based on the second invariant of the velocity gradient tensor and given by
  $$
  Q = \frac12\left(\mathcal R_{ij}\mathcal R_{ij} -\mathcal S_{ij}\mathcal S_{ij}\right),
  $$
  $\mathcal S_{ij}$ is the strain tensor, and $\mathcal R_{ij}$ is the rotation tensor. The region of Q > 0 represents the existence of a vortex. Since POD modes come from two-point correlation tensor, they only captures the statistically representative coherent structures associated with most of the mean energy. In contrast, instantaneous coherent  vortices captured with LES are responsible in transferring turbulence kinetic energy from large to small scales. 

 \section{Conclusion}\label{sec:con}
 In this article, we model the fluid flow around arbitrary solid objects using the direct-forcing immersed boundary method. The results show that the POD representation of the flow is relatively insensitive to the shape of the solid geometry as long as the Reynolds number of the flow is not very large. We find that the POD modes learn the statistically representative coherent flow structure regardless of what is the shape of the obstacle. In other words, accurate ROM of flow around obstacles are derived from the Galerkin projection of the Navier-Stokes equation. However, at high Reynolds number, particularly in the atmospheric boundary layer, where the dominant dynamics is not time periodic, the coherent structure identified by the POD modes differ significantly from what is captured by traditional vortex identification methods. More specifically, the POD analysis of atmospheric boundary layer flow suggests that POD method can be combined with the LES method, which may lead to a more accurate POD assisted turbulence modeling strategy. This work is currently underway.

\section*{Acknowledgments} \nonumber

The author would like to thank Compute Canada for providing computational support through a Resource Allocation Project.

\bibliography{bibrefs}




\end{multicols}

\end{document}